\documentclass[a4paper,11pt]{article}

\usepackage{graphicx,amsmath,amssymb}
\usepackage{pos}
\usepackage{wrapfig}
\usepackage{ragged2e}

\newcommand{\lb}{\label}

\title{Limits of applicability of holographic dual descriptions to QCD: virtuality and coherence }

\ShortTitle{Holographic light-front QCD}

\author*[a]{Guy~F.~de~T\'eramond}

\affiliation[a]{Laboratorio de F\'isica Te\'orica y Computacional, \\ Universidad de Costa Rica, 11501 San Jos\'e, Costa Rica}

\emailAdd{guy.deteramond@ucr.ac.cr}

\abstract{Virtuality and coherence determine the limits of applicability of holographic concepts in QCD. In light-front quantization, the invariant mass controls the off-shell behavior of a physical process and thus provides a measure of its virtuality. In impact space, the corresponding quantity is the invariant transverse separation between quarks and gluons, $\zeta$, which is identified with the holographic coordinate $z$ in holographic light-front QCD. By embedding the internal structure of hadrons and implementing an effective superconformal symmetry, the mapping between quantum states and classical gravity --central to the holographic principle-- yields new analytic insights into the strong-interaction dynamics responsible for the emergence of a confinement scale and the observed hadron spectrum.  We show that the possible emergence of a gravity dual to physical QCD is restricted to the Regge limit of high-energy scattering, where many partonic degrees of freedom participate coherently. This criterion provides a quantitative definition of the domain of applicability of holographic QCD to dynamical processes and allows it to be tested directly against experimental results. By further enforcing analyticity together with exact QCD constraints at asymptotic infinity, the holographic framework can be consistently extended beyond the infrared domain. To illustrate this procedure, we briefly discuss recent work by the HLFHS collaboration, where the effective strong coupling was extended from the Regge-limit domain, through the near-perturbative transition region, and into the ultraviolet Bjorken-limit domain. The resulting scheme provides a unified and precise nonperturbative description of the strong coupling across all virtuality scales.

\vspace{20pt}

\begin{flushright}
{\small \it ``High-energy scattering and deep inelastic processes \\
are two aspects of the same amplitude, \\
connected through virtuality and coherence.'' \\
\rm Attributed to V.\,N.\ Gribov.}
\end{flushright}
}

\FullConference{QCD at the Extremes (QCDEX2025)
\\1-5 September 2025 
\\ Online}

\begin{document}
 
\maketitle

\section{The holographic principle and limits of applicability of dual descriptions to QCD}

According to the holographic principle, the physics of gravity within a spacetime volume can be encoded in a strongly coupled quantum-mechanical system with a large number of degrees of freedom defined on its boundary~\cite{tHooft:1993dmi,Susskind:1994vu}. The AdS/CFT correspondence~\cite{Maldacena:1997re} originates in string theory and thus provides a precise mathematical realization of this principle. It establishes the equivalence between a weakly coupled theory of gravity in a higher-dimensional anti--de Sitter (AdS) space --a spacetime of constant negative curvature-- and a strongly coupled conformal field theory (CFT) with a large number of colors, $N_c$, defined on its lower-dimensional Minkowski boundary.

Quantum Chromodynamics (QCD), a non-Abelian gauge theory with local SU(3) color symmetry, is firmly anchored in physical reality. It is neither supersymmetric nor conformal, nor does it possess a parametrically large value of $N_c$ in the real world. The possible emergence of a gravity dual to physical QCD cannot be inferred from spectroscopy alone. Although a bound state contains an infinite number of higher Fock components beyond the valence configuration, and therefore an infinite number of virtual constituents, this consideration by itself is insufficient to establish the conditions under which a dual description may arise. To assess whether a gravity dual to physical QCD can actually emerge, it is therefore necessary to extend this incomplete notion of confinement beyond the spectroscopic domain to dynamical processes, where the number of active degrees of freedom is explicitly introduced through the interpretation of high-energy experiments.  Thus, the question of whether QCD has a sufficient number of active degrees of freedom for a dual description to be meaningful in a given kinematical domain can, in principle, be answered experimentally.

Early measurements of the proton structure function $F_2(x, Q^2)$ at HERA~\cite{H1:2001ert}, in the kinematic range
\(
5 \times 10^{-5} \le x \le 0.01,
\)
point to a gluon density that increases toward small values of the Bjorken variable $x$, consistent with Regge behavior and Pomeron exchange. Other dynamical processes in the Regge limit, where the center-of-mass energy $s$ is much larger than the momentum transfer $t$, $s \gg |t|$, are also consistent with a high density of gluons at very small
\(x \sim Q^2/s\),
at fixed $t=-Q^2$. These include elastic and diffractive scattering, hadron electroproduction, inclusive and diffractive deep inelastic scattering (DIS), and, more recently, studies of energy--flow correlators from jets at extreme energies (see, for example, Ref.~\cite{Csaki:2025abk}).

The relation between the gluon density and the rise of the total proton--proton cross section validates the dominance of linear Regge trajectories and Pomeron exchange in forward scattering processes over six orders of magnitude in the center-of-mass energy $s$, corresponding to values of $x \sim Q^2/s$ as small as $x \sim 10^{-9}$~\cite{Dosch:2023bxj}. One may also consider a $Q^2$-dependent Pomeron to describe the $Q^2$ evolution of the gluon distribution function $g(x,Q^2)$ at small $x$~\cite{Dosch:2022mop}. This procedure allows one to describe the $Q^2$ dependence of $F_2(x,Q^2)$ observed at HERA, as well as the energy and $Q^2$ dependence of high-energy diffractive processes involving virtual photons up to LHC energies. This simple alternative to a hard Pomeron illustrates that it is not necessary to modify standard Regge-limit Pomeron dynamics in order to extend holographic results to processes at higher $Q^2$.

 \section{Virtuality and coherence in light-front QCD}
 
In relativistic quantum field theory, virtuality, often denoted by $Q^2$, is a measure of how far a particle is off--mass--shell $Q^2 = P^2 - M^2$. In QCD, virtuality has a special significance since the fundamental constituents of hadrons are never on--mass--shell: $p_i^2 \ne m_i^2$, they are not directly observable particles. It is convenient to define virtuality $\epsilon_\kappa(Q^2)$ with respect to a confinement scale $\kappa$, namely 
\begin{align}  \lb{epka}
\epsilon_\kappa(Q^2) \equiv Q^2 + 4 \kappa^2,
\end{align}
with $\epsilon_\kappa(Q^2 = 0) = 4 \kappa^2$, and $4 \kappa^2 \simeq 1 {\rm GeV}^2$ defined in terms of the proton mass, $4 \kappa^2 = M_p^2$, for reasons that will become apparent below. It is also useful to define coherence length $L_\kappa$ relative to the confinement scale $\kappa$ by  
\begin{align} \lb{Lka}
L_\kappa^2(Q^2)  \equiv  {\epsilon_\kappa}^{-1}(Q^2).
\end{align}
At low virtuality states maintain coherence over long light-front times and large spacetime intervals, thus allowing nonperturbative dynamics to form a bound state. At higher virtuality scales, $Q^2 \gg 4 \kappa^2$, virtuality grows with the probe scale,
\(
\epsilon_\kappa(Q^2 \gg 4 \kappa^2) \sim Q^2 ,
\)
and the coherence $L_\kappa(Q^2)$ it is gradually lost. This is the case for DIS in the Bjorken limit, $Q^2 \to \infty$, which revealed the underlying point-like structure of the proton constituents.

In light front (LF) quantization~\cite{Dirac:1949cp} the LF invariant Hamiltonian, $P^2$, for a hadron with four-momentum 
\(
P^\mu = \left(P^+, P^-, \mathbf{P}_\perp \right)
\)
is given by 
\(
P^2 = P^\mu P_\mu = P^+ P^- - \mathbf{P}_{\perp}^{2}.
\)
In the case of the proton, the mass and the confinement scale terms cancel out from the expression of the relative virtuality, $\epsilon_\kappa(Q^2 )$,  with the result
\(
\epsilon_\kappa(Q^2)_p = P^+ P^- + M_n^2,
\)
where  $M_n^2$, the invariant mass, is given by
\begin{align} \lb{Mn}
M_n^2 = \left(\sum_{i=1}^{n} k_i^\mu \right)^2
= \sum_{i=1}^{n} \frac{\mathbf{k}_{\perp i}^{2} + m_i^2}{x_i} ,
\end{align} 
for an $n$-parton Fock state. The longitudinal momentum fraction of each constituent 
\(
x_i = k_i^+ / P^+,
\)
and the transverse momenta $\mathbf{k}_{\perp i}$ are relative variables; \emph{i.e.}, they are independent of the total longitudinal momentum $P^+$ and transverse momentum $\mathbf{P}_{\perp}$ of the hadron bound state, and the  $m_i$ denote the parton masses. The invariant mass~\eqref{Mn} is the key dynamical variable that controls the off-shell behavior of the wave function~\cite{Brodsky:1997de}.

\section{Spectroscopic Regge domain in QCD}  

In QCD the spectroscopic Regge domain in QCD refers to the observed patters of hadron masses and spins obseved in high-energy scattering --the Regge limit-- and often described with holographic models of QCD~\cite{Gross:2022hyw}.  This connection becomes rather precise within the holographic light-front QCD approach  (HLFQCD) in terms of the invariant LF coherence length $L_\kappa$ defined above.

\subsection{Semiclassical approximation to light-front QCD and wave equations in AdS space}

It is useful to consider the conjugate invariant variable to $M_n^2$, Eq.~\eqref{Mn},  in impact space. It is  labeled $\zeta$ and represents the invariant transverse separation between the constituents quark and gluons within a hadron\footnote{For a two-component bound-state $M_{q \bar q} = \mathbf{k^2}_\perp/x(1-x)$ and $\zeta^2 = x(1-x) \mathbf{b}_\perp^2$, with $\mathbf{b}_\perp$ the transverse impact variable conjugate to $\mathbf{k}_\perp$.}. This choice of the variable $\zeta$  allows us to separate kinematics from dynamics in the LF Hamiltonian bound-state problem and to derive a single variable LF quantum mechanical wave equation~\cite{deTeramond:2008ht}.
\begin{align} \lb{LFHWE}
\left(-\frac{d^2}{d\zeta^2} 
- \frac{1 - 4 L^2}{4 \zeta^2}+ U(\zeta) \right)  \phi(\zeta) =  M^2 \phi(\zeta),
\end{align}
where $L_\kappa$ is the LF orbital angular momentum. The effective confinement potential $U(\zeta)$ acts on the valence Fock state, and it is built from the interaction of all the higher Fock components acting coherently over distance  $L_\kappa^2 \sim 1 / \kappa^2$. 

The wave equation in AdS$_{d+1}$  follows from the variation of the action in the presence of a dilaton term, $e^{\varphi(z)}$, responsible for confinement. The dilaton breaks the scaling invariance of the AdS metric and induces an effective interaction $V(z)$. After separating kinematics from dynamics, one obtains a Schr\"odinger-like equation~\cite{deTeramond:2008ht, deTeramond:2013it}
\begin{align} \lb{AdSWE}
\left(-\frac{d^2}{dz^2}  - \frac{1 - 4 \nu^2}{4 z^2}+ V(z) \right)  \Phi(z) = P^2  \, \Phi(z),
\end{align} 
where $z$ is the fifth dimension of AdS$_5$  and $$V(z) =  e^{-\varphi(z)/2} z^{-\gamma - 1}  \partial_z \left(z^{\gamma +1} \, \partial_z e^{\varphi(z)/2}\right),  \quad \gamma = 2J - d,$$ the gravitational potential~\cite{deTeramond:2023qbo}.  Analogous equations are obtained for fermions using Rarita--Schwinger spinors~\cite{deTeramond:2013it}.
 
Equation~\eqref{LFHWE} is a frame-independent relativistic wave equation with a structure similar to that of Eq.~\eqref{AdSWE}. It leads to an exact mapping of the AdS equations to Hamiltonian LF equations, provided that the holographic variable $z$ is identified with the LF invariant transverse distance $\zeta$, $z=\zeta$. Thus, small values of $\zeta$ correspond to large-virtuality processes and to values of $z$ near the AdS boundary. Conversely,  large values of $\zeta$ describe low-virtuality configurations and correspond to large $z$, deep in AdS space, where the metric is modified to introduce nonperturbative dynamics. In addition, the AdS parameter $\nu$ is identified with the orbital angular momentum in the transverse LF plane $L$; The internal structure of hadrons and their angular momentum become essential elements of the holographic light front QCD (HLFQCD) framework~\cite{Brodsky:2014yha}. Finally, we identify the AdS wave function $\Phi$ with the LF eigenfunction $\phi$; the gravitational potential $V$ with the confinement potential $U$; and the Lorentz invariant $P^2=P_\mu P^\mu$ with its eigenvalue $M^2$, the hadron bound-state mass.

 \subsection{Introduction of a mass scale and superconformal symmetry}

The introduction of a mass scale --absent by construction in the classically scale-invariant QCD Lagrangian-- implies the emergence of an effective, nonlocal dynamical symmetry acting over distances of order~$L_\kappa^2 \simeq 1/\kappa^2$. Superconformal symmetry provides a natural realization of this mechanism: the emergence of a mass scale through a graded superconformal algebra uniquely fixes the form of the effective confining potential~\cite{deAlfaro:1976vlx, Fubini:1984hf}, leading to the observed hadronic spectra and to supersymmetric relations among the Regge trajectories of mesons and baryons. Within this framework, the pion is massless in the chiral limit and has a special role, being responsible for the breaking of hadronic supersymmetry~\cite{Dosch:2015nwa, deTeramond:2014asa}.  Extensions of this approach to tetraquarks, as well as to bound states containing heavy quarks, have also been studied~\cite{Brodsky:2016yod, Dosch:2015bca, Dosch:2016zdv, Nielsen:2018ytt}.

The derivation of a unique confinement potential from an effective superconformal symmetry leads to the meson (M) and baryon (B) spectrum~\cite{Dosch:2015nwa}
\begin{align}
\lb{Mmass}
M_M^2 &= 4 \lambda \left(n + L_M\right) , \\
\lb{Bmass}
M_B^2 &= 4 \lambda \left(n + L_B + 1\right)  ,
\end{align}
where $n$ and $L$ label radial and orbital excitations.  Comparing Eqs.~\eqref{Mmass} and~\eqref{Bmass}, one finds the superconformal relation $L_M = L_B + 1$ within a meson--baryon multiplet.  The prediction of a massless meson for $L_M = 0$ (Eq.~\eqref{Mmass}) implies $L_B = -1$, thus signaling the breaking of hadronic supersymmetry in the chiral limit. This pattern holds across all particle families, including hadrons containing heavy quarks. The mass correction terms from heavy quarks, as well as spin-dependent terms, should be included in the longitudinal component of the total LF Hamiltonian, $H = H_\parallel+ H_\perp$, (see, for example, Ref.~\cite{deTeramond:2021yyi}), as required by the superconformal symmetry which acts only in the LF transverse components.

Mesons and baryons lie on linear Regge trajectories, $\alpha(t)=\alpha(0)+\alpha' t$, with $t=M^{2}$, whenever $\alpha(t)$ assumes non-negative integer and half-integer values $J$, such that $\alpha(t=M^{2})=J$. The Regge slope,
\(
\alpha' = \frac{1}{4\lambda},
\)
is universal, reflecting the underlying superconformal structure, whereas the intercept $\alpha(0)$, differs for mesons and baryons~\cite{Dosch:2025lwb}. If we identify the confinement strength $\kappa$ with the mass scale $\lambda$, Eqs.~\eqref{Mmass} and~\eqref{Bmass}, we find an specific relation between the coherence length $L_\kappa$, in the limit $Q^2 \to 0$, and the universal Regge slope
\begin{align} \lb{kalpL}
L_\kappa \equiv L_\kappa^2(Q^2 \to 0)  = \alpha' = \frac{1}{4 \kappa^2}.
\end{align}
It thus corresponds to the maximal coherence length, where an infinite number of quark and gluon degrees of freedom, above the valence Fock state, act collectively. From Eq.~\eqref{Bmass} the proton mass is given by $M_p^2 = 4 \kappa^2 \simeq 1 {\rm GeV}^2$, thus $\kappa^2 \simeq 0.5 ~ {\rm GeV}^2$. Since the Regge slope is universal, the coherence length $L_\kappa$ is also universal for light hadrons, but it becomes progressively smaller for heavy hadron spectrum, since $\alpha'$ increases as a function of the heavy quark mass scale of the bound state~\cite{Branz:2010ub, Dosch:2016zdv, Nielsen:2018ytt}.

The superconformal effective symmetry in HLFQCD predicts identical values of the confinement scale $\kappa$ for mesons and baryons, but leaves its absolute magnitude undetermined. Indeed, the hadronic mass scale cannot be derived from first-principles QCD using either perturbative or nonperturbative methods: it is not contained within the intrinsic structure of the theory and therefore requires an external input, such as the measured proton mass
\(
4\kappa^{2} = M_p^{2},
\)
or the Regge slope $\alpha'$. Both are connected to the coherence length $L_\kappa$ by Eq.~\eqref{kalpL}.

\section{Ultraviolet completion of analytic amplitudes in QCD}

The introduction of Veneziano duality from the pre-QCD era~\cite{Veneziano:1968yb} is an important element in the computation of form factors and parton distribution functions. A straightforward application of the holographic QCD framework, for example using the soft-wall model~\cite{Karch:2006pv}, leads to incorrect results, since it does not incorporate the Regge intercept, which is essential for a meaningful comparison with experimental observations. Duality therefore becomes a key ingredient at the level of scattering amplitudes and form factors.

By imposing strict boundary conditions in the deep ultraviolet limit, the framework can be extended to describe hadronic form factors as well as quark and gluon distributions~\cite{deTeramond:2018ecg,deTeramond:2021lxc}. As recently shown by the HLFHS collaboration, enforcing analyticity together with exact QCD constraints at asymptotic infinity also makes it possible to extend the effective strong coupling from the infrared holographic regime into the ultraviolet domain~\cite{deTeramond:2025qlj}.

\subsection{All scales effective running coupling in QCD} 

The effective strong coupling $\alpha_{\rm eff}$ given in~\cite{deTeramond:2024ikl} 
\begin{align}\lb{gen}\
\alpha_{\rm eff}(Q^2) = \alpha_{\rm eff}(0) \exp\Big[-\int_0^{Q^2} \frac{du} {4 \kappa^2 +u \log(\frac{u}{\Lambda^2})}\Big],
\end{align}
behaves in the Regge-limit domain and the Bjorken-limit region as follows:
\begin{align}
\alpha_{\rm eff}(Q^2) \to \left\{
\begin{array}{lcl}
e^{-Q^2/4\kappa^2}, &\mbox{for } Q^2\ll 4 \kappa^2 , \\
\frac{1}{\log(Q^2/\Lambda^2)}, &\mbox{for } Q^2\gg 4 \kappa^2 .
\end{array}\right.
\end{align}
For $Q^2\gg  4 \kappa^2$ this is the typical leading order perturbative behavior. For $Q^2\ll 4 \kappa^2$, it is the non-perturbative behavior based on the light front holographic approach~\cite{Brodsky:2010ur, deTeramond:2024ikl, deTeramond:2025qlj}.

\begin{figure}
\centering
\includegraphics[width=7.6cm]{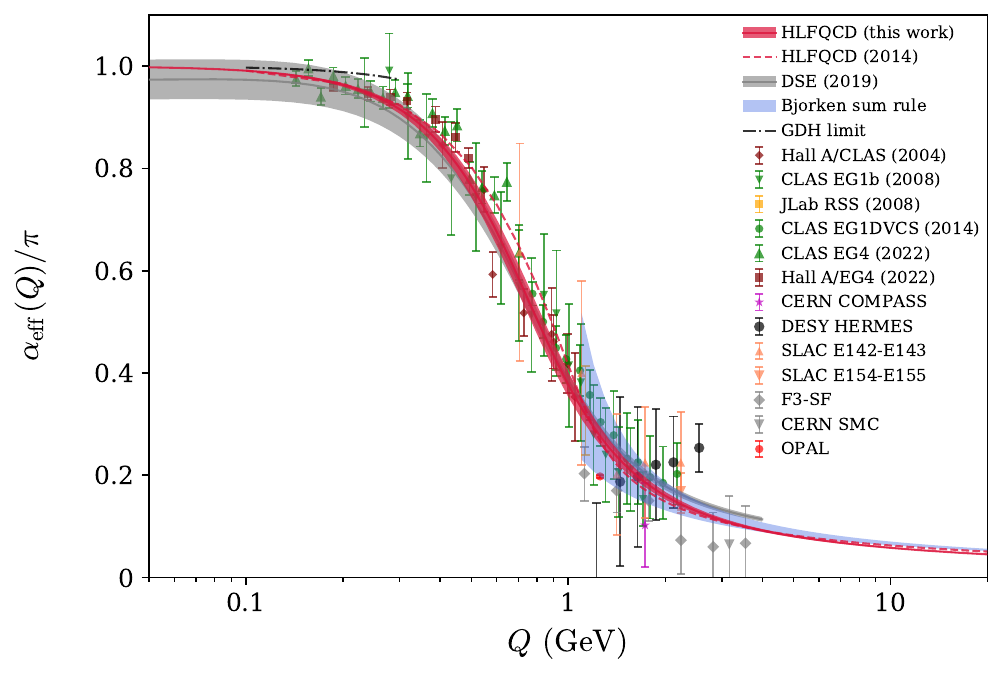}  
\includegraphics[width=7.4cm]{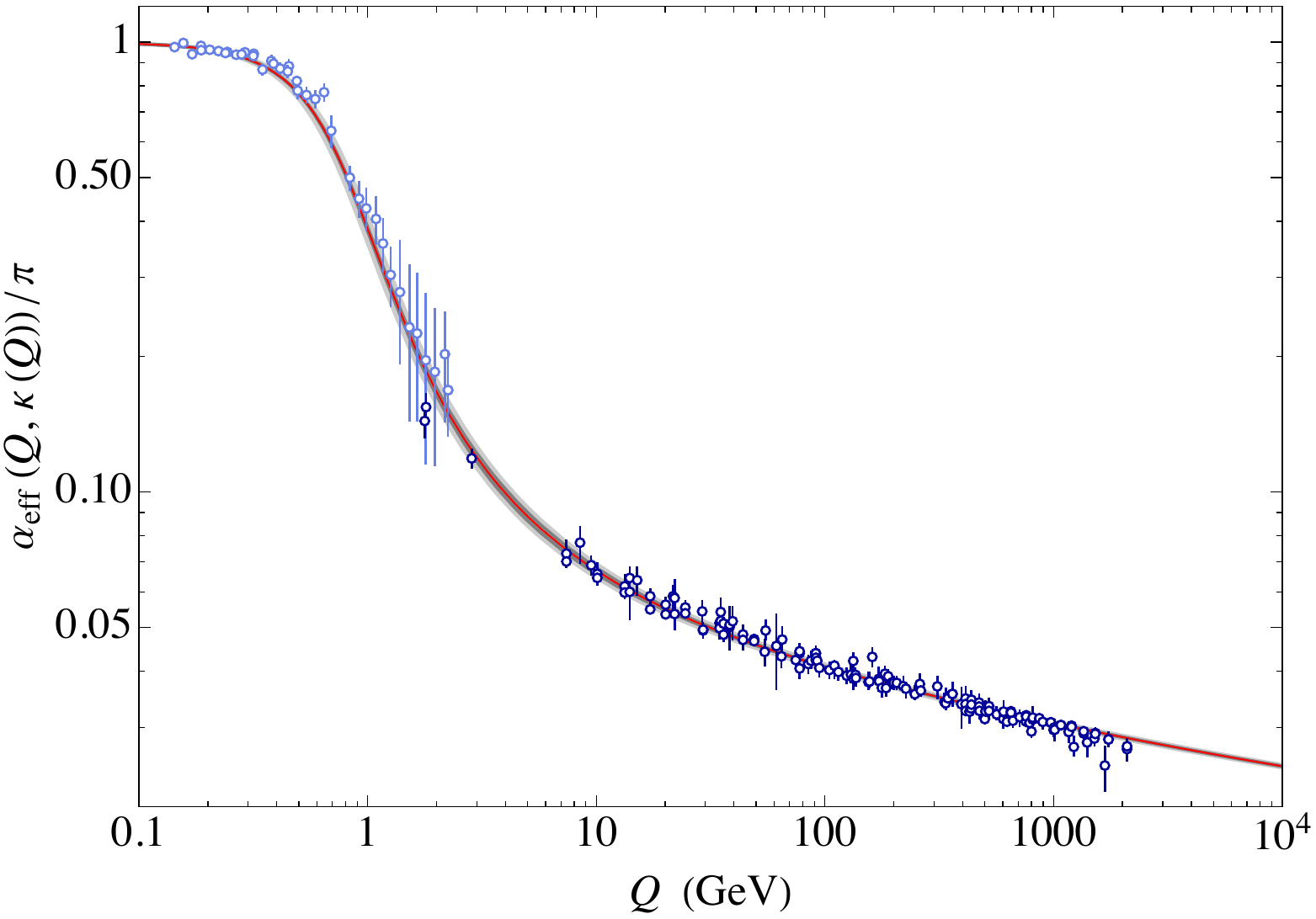}
\caption{\lb{Five} The effective coupling $\alpha_{\rm eff}(Q)$. The results shown in the figure correspond to $\kappa = 0.534~\rm{GeV}$.  (left) The IR and nonperturbative transition domain, (right) imposing UV completion and incorporating the influence of heavy quark thresholds
~\cite{deTeramond:2025qlj}.}
\end{figure}

In Ref.~\cite{deTeramond:2024ikl}, it was shown that analyticity arguments allows one to determine the perturbative constant $\Lambda$ in terms of of the confinement scale as $\kappa$ as $\Lambda= \frac{8}{\pi} \kappa^2$. Thus, the transition from the hadronic to the perturbative region, approximately the region $2 \leq Q \leq 5$ GeV, is also governed by the confinement scale $\kappa$, as can be seen from Fig.~\ref{Five} (left).  

Above this region, the influence of heavy quarks, as well as the QCD boundary conditions in the asymptotic limit $Q^2 \to \infty$, are critical elements of the model. Threshold effects are incorporated while preserving the analytic properties and the UV completion require to consider the evolution of the confinement scale $\kappa$ following the substitution $\kappa^2 \to \kappa^2 f(u) \equiv \kappa(u)$ with $f(u)$ a scaling factor.  The scaling properties are compatible with the introduction of branch cuts in the time-like domain of the complex plane, with no overlapping singularities in the space-like domain, therefore preserving the flow of singularities and the maximal analyticity properties in the space-like domain found in~\cite{deTeramond:2024ikl}. The model requires the introduction of three heavy quark weight coefficients $C_q$, which are strongly constrained by asymptotic sum rules. As a result, one obtains the all-scales result shown in Fig.~\ref{Five} (right).

\section{Concluding Remarks and Outlook}

We have presented an overview of QCD dynamics across energy scales from the perspective of light-front holography, analyticity, and QCD ultraviolet boundary conditions. Describing physics over such vastly different regimes requires relating complementary methods and concepts, ranging from the coherent, low-virtuality Regge-limit domain --where holographic QCD provides meaningful results-- to the incoherent, high-virtuality Bjorken-limit domain of deep inelastic scattering, where perturbative QCD applies successfully. Bridging these two domains requires enforcing analyticity and ultraviolet completion in physical amplitudes, which is essential for a meaningful comparison with experimental observations.  Within this framework, Regge theory, together with pre-QCD concepts such as Veneziano duality, becomes increasingly intertwined with fundamental QCD. The introduction of an effective superconformal symmetry requires the emergence of a mass scale in the theory and uniquely fixes the confining potential. These key elements are not imposed phenomenologically, but instead follow from the effective symmetries of the model.

We have shown how these concepts can be applied to a specific problem, namely the nonperturbative description of the effective QCD coupling across all scales. Similar considerations have been applied successfully to the description of hadron form factors and quark and gluon distribution functions, and will be discussed elsewhere along the lines developed in this article.

The theoretical inevitability of introducing a mass scale starting from the QCD Lagrangian, however, remains an open question. From an analytic point of view, QCD is directly applicable at very large $Q^2$, where nonperturbative elements --such as the mass scale-- are decoupled at order $ 4\kappa^2/Q^2$. Sensible answers must therefore take into account the analytic bridge between these two domains, providing a coherent analytic understanding of nonperturbative phenomena in the absence of a formal proof of confinement.

\textbf{Acknowledgements}.  
 I am grateful to Stanley Brodsky and Hans Guenter Dosch for their invaluable collaboration in the development of light-front holography, and to Alexandre Deur, Tianbo Liu, Arpon Paul, and Raza Sabbir Sufian for their significant contributions to the recent applications of holographic light-front QCD. I thank the organizers of the 2025 workshop \emph{QCD at the Extremes}, H.~Jung, K.~Kutak, N.~Raicevic, and S.~Taheri-Monfared, for their kind invitation, and Alexandre Deur for insightful discussions on $\alpha_s$.

%%%%%%%%%%%%%%%%%%
%\bibliographystyle{JHEP}
%\bibliography{HLFQCD}

\begin{thebibliography}{99}

\bibitem{tHooft:1993dmi}
G.~'t~Hooft, \emph{{Dimensional reduction in quantum gravity}}, {\emph{Conf.
  Proc. C} {\bfseries 930308} (1993) 284}
  [\href{https://arxiv.org/abs/gr-qc/9310026}{{\ttfamily gr-qc/9310026}}].

\bibitem{Susskind:1994vu}
L.~Susskind, \emph{{The world as a hologram}},
  \href{https://doi.org/10.1063/1.531249}{\emph{J. Math. Phys.} {\bfseries 36}
  (1995) 6377} [\href{https://arxiv.org/abs/hep-th/9409089}{{\ttfamily
  hep-th/9409089}}].

\bibitem{Maldacena:1997re}
J.M.~Maldacena, \emph{{The large $N$ limit of superconformal field theories and
  supergravity}}, \href{https://doi.org/10.4310/ATMP.1998.v2.n2.a1}{\emph{Adv.
  Theor. Math. Phys.} {\bfseries 2} (1998) 231}
  [\href{https://arxiv.org/abs/hep-th/9711200}{{\ttfamily hep-th/9711200}}].

\bibitem{H1:2001ert}
{\scshape H1} collaboration, \emph{{On the rise of the proton structure
  function $F_2$ towards low $x$}},
  \href{https://doi.org/10.1016/S0370-2693(01)01074-7}{\emph{Phys. Lett. B}
  {\bfseries 520} (2001) 183}
  [\href{https://arxiv.org/abs/hep-ex/0108035}{{\ttfamily hep-ex/0108035}}].

\bibitem{Csaki:2025abk}
C.~Cs{\'a}ki, A.~Ismail and L.~Kiriliuk, \emph{{Asymptotic freedom for
  holographic energy correlators}},
  \href{https://arxiv.org/abs/2511.03778}{{\ttfamily 2511.03778}}.

\bibitem{Dosch:2023bxj}
H.G.~Dosch, G.F.~de~Teramond and S.J.~Brodsky, \emph{{Entropy from entangled
  parton states and high-energy scattering behavior}},
  \href{https://doi.org/10.1016/j.physletb.2024.138521}{\emph{Phys. Lett. B}
  {\bfseries 850} (2024) 138521}
  [\href{https://arxiv.org/abs/2304.14207}{{\ttfamily 2304.14207}}].

\bibitem{Dosch:2022mop}
{\scshape HLFHS} collaboration, \emph{{Towards a single scale-dependent Pomeron
  in holographic light-front QCD}},
  \href{https://doi.org/10.1103/PhysRevD.105.034029}{\emph{Phys. Rev. D}
  {\bfseries 105} (2022) 034029}
  [\href{https://arxiv.org/abs/2201.09813}{{\ttfamily 2201.09813}}].

\bibitem{Dirac:1949cp}
P.A.M.~Dirac, \emph{{Forms of relativistic dynamics}},
  \href{https://doi.org/10.1103/RevModPhys.21.392}{\emph{Rev. Mod. Phys.}
  {\bfseries 21} (1949) 392}.

\bibitem{Brodsky:1997de}
S.J.~Brodsky, H.-C.~Pauli and S.S.~Pinsky, \emph{{Quantum chromodynamics and
  other field theories on the light cone}},
  \href{https://doi.org/10.1016/S0370-1573(97)00089-6}{\emph{Phys. Rept.}
  {\bfseries 301} (1998) 299}
  [\href{https://arxiv.org/abs/hep-ph/9705477}{{\ttfamily hep-ph/9705477}}].

\bibitem{Gross:2022hyw}
F.~Gross et~al., \emph{{50 Years of quantum chromodynamics}},
  \href{https://doi.org/10.1140/epjc/s10052-023-11949-2}{\emph{Eur. Phys. J. C}
  {\bfseries 83} (2023) 1125}
  [\href{https://arxiv.org/abs/2212.11107}{{\ttfamily 2212.11107}}].

\bibitem{deTeramond:2008ht}
G.F.~de~T\'eramond and S.J.~Brodsky, \emph{{Light-front holography: A first
  approximation to QCD}},
  \href{https://doi.org/10.1103/PhysRevLett.102.081601}{\emph{Phys. Rev. Lett.}
  {\bfseries 102} (2009) 081601}
  [\href{https://arxiv.org/abs/0809.4899}{{\ttfamily 0809.4899}}].

\bibitem{deTeramond:2013it}
G.F.~de~T\'eramond, H.G.~Dosch and S.J.~Brodsky, \emph{{Kinematical and
  dynamical aspects of higher-spin bound-state equations in holographic QCD}},
  \href{https://doi.org/10.1103/PhysRevD.87.075005}{\emph{Phys. Rev. D}
  {\bfseries 87} (2013) 075005}
  [\href{https://arxiv.org/abs/1301.1651}{{\ttfamily 1301.1651}}].

\bibitem{deTeramond:2023qbo}
G.F.~de~T\'eramond and S.J.~Brodsky, \emph{{Color symmetry and confinement as
  an underlying superconformal structure in holographic QCD}},
  \href{https://doi.org/10.1142/S0217751X24410070}{\emph{Int. J. Mod. Phys. A}
  {\bfseries 39} (2024) 2441007}
  [\href{https://arxiv.org/abs/2308.09280}{{\ttfamily 2308.09280}}].

\bibitem{Brodsky:2014yha}
S.J.~Brodsky, G.F.~de~T\'eramond, H.G.~Dosch and J.~Erlich, \emph{{Light-front
  holographic QCD and emerging confinement}},
  \href{https://doi.org/10.1016/j.physrep.2015.05.001}{\emph{Phys. Rept.}
  {\bfseries 584} (2015) 1} [\href{https://arxiv.org/abs/1407.8131}{{\ttfamily
  1407.8131}}].

\bibitem{deAlfaro:1976vlx}
V.~de~Alfaro, S.~Fubini and G.~Furlan, \emph{{Conformal invariance in quantum
  mechanics}}, \href{https://doi.org/10.1007/BF02785666}{\emph{Nuovo Cim. A}
  {\bfseries 34} (1976) 569}.

\bibitem{Fubini:1984hf}
S.~Fubini and E.~Rabinovici, \emph{{Superconformal quantum mechanics}},
  \href{https://doi.org/10.1016/0550-3213(84)90422-X}{\emph{Nucl. Phys. B}
  {\bfseries 245} (1984) 17}.

\bibitem{Dosch:2015nwa}
H.G.~Dosch, G.F.~de~T\'eramond and S.J.~Brodsky, \emph{{Superconformal
  baryon-meson symmetry and light-front holographic QCD}},
  \href{https://doi.org/10.1103/PhysRevD.91.085016}{\emph{Phys. Rev. D}
  {\bfseries 91} (2015) 085016}
  [\href{https://arxiv.org/abs/1501.00959}{{\ttfamily 1501.00959}}].

\bibitem{deTeramond:2014asa}
G.F.~de~T\'eramond, H.G.~Dosch and S.J.~Brodsky, \emph{{Baryon spectrum from
  superconformal quantum mechanics and its light-front holographic embedding}},
  \href{https://doi.org/10.1103/PhysRevD.91.045040}{\emph{Phys. Rev. D}
  {\bfseries 91} (2015) 045040}
  [\href{https://arxiv.org/abs/1411.5243}{{\ttfamily 1411.5243}}].

\bibitem{Brodsky:2016yod}
S.J.~Brodsky, G.F.~de~T{\'e}ramond, H.G.~Dosch and C.~Lorc{\'e},
  \emph{{Universal effective hadron dynamics from superconformal algebra}},
  \href{https://doi.org/10.1016/j.physletb.2016.05.068}{\emph{Phys. Lett. B}
  {\bfseries 759} (2016) 171}
  [\href{https://arxiv.org/abs/1604.06746}{{\ttfamily 1604.06746}}].

\bibitem{Dosch:2015bca}
H.G.~Dosch, G.F.~de~T\'eramond and S.J.~Brodsky, \emph{{Supersymmetry across
  the light and heavy-light hadronic spectrum}},
  \href{https://doi.org/10.1103/PhysRevD.92.074010}{\emph{Phys. Rev. D}
  {\bfseries 92} (2015) 074010}
  [\href{https://arxiv.org/abs/1504.05112}{{\ttfamily 1504.05112}}].

\bibitem{Dosch:2016zdv}
H.G.~Dosch, G.F.~de~T\'eramond and S.J.~Brodsky, \emph{{Supersymmetry across
  the light and heavy-light hadronic spectrum II}},
  \href{https://doi.org/10.1103/PhysRevD.95.034016}{\emph{Phys. Rev. D}
  {\bfseries 95} (2017) 034016}
  [\href{https://arxiv.org/abs/1612.02370}{{\ttfamily 1612.02370}}].

\bibitem{Nielsen:2018ytt}
M.~Nielsen, S.J.~Brodsky, G.F.~de~T\'eramond, H.G.~Dosch, F.S.~Navarra and
  L.~Zou, \emph{{Supersymmetry in the double-heavy hadronic spectrum}},
  \href{https://doi.org/10.1103/PhysRevD.98.034002}{\emph{Phys. Rev. D}
  {\bfseries 98} (2018) 034002}
  [\href{https://arxiv.org/abs/1805.11567}{{\ttfamily 1805.11567}}].

\bibitem{deTeramond:2021yyi}
G.F.~de~T\'eramond and S.J.~Brodsky, \emph{{Longitudinal dynamics and chiral
  symmetry breaking in holographic light-front QCD}},
  \href{https://doi.org/10.1103/PhysRevD.104.116009}{\emph{Phys. Rev. D}
  {\bfseries 104} (2021) 116009}
  [\href{https://arxiv.org/abs/2103.10950}{{\ttfamily 2103.10950}}].

\bibitem{Dosch:2025lwb}
H.G.~Dosch, G.F.~de~T\'eramond and S.J.~Brodsky, \emph{{Holographic light-front
  QCD}},  10, 2025 [\href{https://arxiv.org/abs/2510.20180}{{\ttfamily
  2510.20180}}].

\bibitem{Branz:2010ub}
T.~Branz, T.~Gutsche, V.E.~Lyubovitskij, I.~Schmidt and A.~Vega, \emph{{Light
  and heavy mesons in a soft-wall holographic approach}},
  \href{https://doi.org/10.1103/PhysRevD.82.074022}{\emph{Phys. Rev. D}
  {\bfseries 82} (2010) 074022}
  [\href{https://arxiv.org/abs/1008.0268}{{\ttfamily 1008.0268}}].

\bibitem{Veneziano:1968yb}
G.~Veneziano, \emph{{Construction of a crossing-symmetric, Regge-behaved
  amplitude for linearly rising trajectories}},
  \href{https://doi.org/10.1007/BF02824451}{\emph{Nuovo Cim. A} {\bfseries 57}
  (1968) 190}.

\bibitem{Karch:2006pv}
A.~Karch, E.~Katz, D.T.~Son and M.A.~Stephanov, \emph{{Linear confinement and
  AdS/QCD}}, \href{https://doi.org/10.1103/PhysRevD.74.015005}{\emph{Phys. Rev.
  D} {\bfseries 74} (2006) 015005}
  [\href{https://arxiv.org/abs/hep-ph/0602229}{{\ttfamily hep-ph/0602229}}].

\bibitem{deTeramond:2018ecg}
{\scshape HLFHS} collaboration, \emph{{Universality of generalized parton
  distributions in light-front holographic QCD}},
  \href{https://doi.org/10.1103/PhysRevLett.120.182001}{\emph{Phys. Rev. Lett.}
  {\bfseries 120} (2018) 182001}
  [\href{https://arxiv.org/abs/1801.09154}{{\ttfamily 1801.09154}}].

\bibitem{deTeramond:2021lxc}
{\scshape HLFHS} collaboration, \emph{{Gluon matter distribution in the proton
  and pion from extended holographic light-front QCD}},
  \href{https://doi.org/10.1103/PhysRevD.104.114005}{\emph{Phys. Rev. D}
  {\bfseries 104} (2021) 114005}
  [\href{https://arxiv.org/abs/2107.01231}{{\ttfamily 2107.01231}}].

\bibitem{deTeramond:2025qlj}
{\scshape HLFHS} collaboration, \emph{{Asymptotic gauge symmetry and UV
  extension of the nonperturbative coupling in holographic QCD}},
  \href{https://doi.org/10.1103/tdyb-7ddp}{\emph{Phys. Rev. D} {\bfseries 112}
  (2025) 094010} [\href{https://arxiv.org/abs/2505.19545}{{\ttfamily
  2505.19545}}].

\bibitem{deTeramond:2024ikl}
{\scshape HLFHS} collaboration, \emph{{QCD running coupling in the
  nonperturbative and near-perturbative regimes}},
  \href{https://doi.org/10.1103/PhysRevLett.133.181901}{\emph{Phys. Rev. Lett.}
  {\bfseries 133} 181901} [\href{https://arxiv.org/abs/2403.16126}{{\ttfamily
  2403.16126}}].

\bibitem{Brodsky:2010ur}
S.J.~Brodsky, G.F.~de~T\'eramond and A.~Deur, \emph{{Nonperturbative QCD
  coupling and its $\beta$-function from light-front holography}},
  \href{https://doi.org/10.1103/PhysRevD.81.096010}{\emph{Phys. Rev. D}
  {\bfseries 81} (2010) 096010}
  [\href{https://arxiv.org/abs/1002.3948}{{\ttfamily 1002.3948}}].

\end{thebibliography}
%%%%%%%%%%%%%%%%%%

%%%%%%%%%%%%%%%%%%
\end{document}